\documentclass[12pt, reqno]{article}
\usepackage{authblk}
\usepackage[margin=1in]{geometry}
\RequirePackage{amsthm,amsmath,amsfonts,amssymb}
\usepackage{array}
\RequirePackage{natbib}
\RequirePackage[colorlinks,citecolor=blue,urlcolor=blue]{hyperref}
\usepackage{booktabs,longtable}
\usepackage{tabularx}
\usepackage{multirow}
\usepackage{threeparttable}
\usepackage{graphicx}
\graphicspath{{./}{../image/}}
\usepackage{xcolor}
\usepackage{setspace}
\usepackage{soul}
\usepackage{epstopdf}
\usepackage[inline]{enumitem}
\usepackage{bm}
\usepackage[ruled, lined]{algorithm2e}
\usepackage{rotating}

\allowdisplaybreaks

\newcommand\pkg[1]{\texttt{#1}}
\let\proglang=\textsf 

\newcommand{\bmB}{\bm{B}}
\newcommand{\bmI}{\bm{I}}
\newcommand{\bmS}{\bm{S}}

\newcommand{\bmX}{\bm{X}}

\newcommand{\bmmu}{\bm{\mu}}
\newcommand{\bmSigma}{\bm{\Sigma}}
\newcommand{\bmOmega}{\bm{\Omega}}
\newcommand{\bmV}{\bm{V}}

\newcommand{\hatbmOmega}{\hat{\bm{\Omega}}}

\newcommand{\tildebmOmega}{\tilde{\bm{\Omega}}}
\DeclareMathOperator*{\argmin}{arg \, min}

\newcommand{\tr}{\text{tr}}

\title{Gaussian Graphical Models for Functional Connectivity 
Analysis: A Statistical Review with Applications to Alzheimer's 
Disease}

\author[1,2,3,*]{Panpan Zhang}

\author[4]{Shiying Xiao}

\author[5]{W.\ Hudson Robb}

\author[1,2,3]{Dandan Liu}

\author[2,3,6,7]{Angela L.\ Jefferson}

\author[4]{Jun Yan}

\affil[1]{Department of Biostatistics, Vanderbilt University Medical 
	Center, Nashville, TN 37203, USA}

\affil[2]{Vanderbilt Memory \& Alzheimer's Center, Vanderbilt 
University Medical Center, Nashville, TN 37203, USA}

\affil[3]{Vanderbilt Alzheimer's Disease Research Center, Vanderbilt 
University Medical Center, Nashville, TN 37203, USA}

\affil[4]{Department of Statistics, University of Connecticut, 
Storrs, CT 06269, USA}

\affil[5]{Department of Radiology, Mayo Clinic Arizona, Scottsdale, 
AZ 85259, USA}

\affil[6]{Department of Neurology, Vanderbilt University Medical 
	Center, Nashville, TN 37232, USA}

\affil[7]{Department of Medicine, Vanderbilt University Medical 
	Center, Nashville, TN 37232, USA}

\affil[*]{Correspondence: \href{mailto:panpan.zhang@vumc.org}{Panpan 
Zhang}}

\begin{document}
	
\maketitle
	
\begin{abstract}

Functional connectivity analysis is an important tool for 
characterizing interactions among brain regions, particularly in 
studies of neurodegenerative disorders such as Alzheimer’s disease 
(AD). Gaussian graphical models (GGMs) provide a promising 
statistical framework for estimating functional connectivity by 
capturing conditional dependence relationships among brain regions. 
Although a variety of regularized precision matrix estimators have 
been proposed to estimate sparse conditional dependency structures 
for GGMs, their comparative performance and practical 
implications for neuroimaging studies are not well understood. In 
this work, we present a comprehensive statistical review and 
empirical evaluation of widely used GGM estimation methods, including 
the graphical lasso (glasso), ridge-based glasso, graphical elastic 
net, adaptive glasso, smoothly clipped absolute deviation (SCAD), 
minimax concave penalty (MCP), constrained $\ell_1$ minimization for 
inverse matrix estimation (CLIME), and tuning-insensitive graph 
estimation and regression (TIGER). Their performance is evaluated 
through extensive data-driven simulations designed to reflect 
realistic neuroimaging settings, along with an application to an AD 
cohort study to illustrate methodological differences and their 
impact on downstream network analysis. In addition, a user-friendly 
\proglang{R} package, \pkg{spice}, is provided to facilitate 
implementation and enhance the reproducibility of empirical 
studies.


\end{abstract}

\doublespacing

\section{Introduction}
\label{sec:intro}

Functional magnetic resonance imaging (fMRI) data provide insight
into intrinsic interactions among 
brain regions and the pathogenesis of neurological disease, such as 
Alzheimer's disease~\citep[AD,][]{dickerson2009large, 
dennis2014functional, franzmeier2017functional}. This non-invasive 
technique detects functional changes in brain activity and 
connectivity that often precede structural alterations in 
AD~\citep{sperling2011thepotential, damoiseaux2012resting}. As such,
it holds potential as a biomarker for detecting AD and monitoring 
AD progression~\citep{sorg2007selective, brier2012loss, 
sheline2013resting}. Graph-theory-based network analysis 
methods~\citep{bullmore2009complex, sporns2018graph} enable
exploring functional changes in brains. These methods
enhance understanding of connection structures by identifying
abnormal regional spontaneous fluctuation of blood oxygenation 
level-dependent (BOLD) indications and altered functional
connectivity patterns linked to AD~\citep{supekar2008network,  
rubinov2010complex, filippi2011structural}.

Brain networks, where nodes represent regions of 
interest (ROIs) across the whole brain or a specific functional 
region (e.g., default mode network, somatomotor network or dorsal 
attention network), form the basis of functional connectivity 
analysis. In such networks, (weighted) links  embody estimated 
functional connectivity among ROIs. A common approach to estimating 
this connectivity is using the Pearson correlation of time series 
extracted from brain regions. However, Pearson correlation is 
sensitive to confounding factors, often leading to spurious 
connections in functional brain networks~\citep{bastos2016atutorial, 
wang2016anefficient}, and potentially resulting in inaccurate 
inference and misleading conclusion in downstream 
analyses~\citep{reid2019advancing}. To address this issue, partial 
correlation has been employed as it estimates functional
connectivity while accounting for potential 
confounders~\citep{marrelec2006partial, ryali2012estimation, 
wang2016anefficient}, offering a more reliable method for
constructing functional brain networks in neuroimaging 
studies~\citep{qiu2016joint, warnick2018abayesian, 
lukemire2021bayesian}.

Partial correlation estimation is fundamentally linked to precision
matrix estimation, a core component of Gaussian Graphical
models~\citep[GGMs,][]{lauritzen1996graphical}. Estimating the 
precision matrix is computationally challenging, particularly under 
high dimensional settings, i.e., the matrix dimension is much higher 
than sample size. In computational statistics, various methods have 
been proposed to tackle this problem~\citep{guo2011joint, 
leng2012sparse, chen2013covariance, banerjee2015bayesian, 
liu2015fast, tsai2022joint}. To date, there has been a lack of 
comprehensive evaluation of precision matrix estimation methods 
specifically for functional connectivity analyses. Even for graphical 
lasso (\underline{l}east 
\underline{a}bsolute \underline{s}hrinkage and \underline{s}election 
\underline{o}perator), one of the most widely used approaches in 
neuroimaging research~\citep{belilovsky2016testing, 
ryali2012estimation, liegeois2020revisiting}, systematic assessment 
has been limited.

This paper addresses this gap through four contributions 
to graph-based functional connectivity analysis. First, it offers a 
comprehensive review of precision matrix estimation methods, 
emphasizing partial correlation in high-dimensional fMRI data. 
Second, it conducts a systematic empirical comparison of
representative approaches through extensive simulations and 
applications to real-world data from the Tennessee Alzheimer’s 
Project (TAP).  By linking methodological comparison with real-data 
analysis, the study shows how estimator choice influences inferred 
functional connectivity patterns and their interpretation in AD 
research. Third, it reviews available \proglang{R} package 
implementations, providing practical guidance
on tuning parameter selection and computational considerations in
functional connectivity analysis. Finally, to facilitate 
implementation and reproducibility, we introduce a
unified package, \pkg{spice}~\citep{pkg:spice}, which provides a 
consistent interface to existing implementations and dispatches to 
underlying packages within a standardized workflow.

The rest of the manuscript is organized as follows. 
Section~\ref{sec:pre} introduces key notations that will be 
used throughout the paper, and briefly reviews the fundamentals 
of partial correlation and GGM. Section~\ref{sec:method} details 
each considered method, including its analytical form, available 
software packages, and current applications in functional 
connectivity research. Section~\ref{sec:sim} 
presents simulation results, along with method assessments and 
sensitivity analyses. Section~\ref{sec:app} explores practical 
applications to the TAP data. Finally, Section~\ref{sec:dis} 
provides a summary of findings and discusses directions for future 
research.

\section{Notations and Preliminaries}
\label{sec:pre}

This section introduces the notations that will be used throughout 
the paper as well as foundational concepts for GGMs
in the context of functional brain network 
analysis. Let $G := G(V, E)$ denote a functional brain network, where 
$V$ is the set of nodes representing ROIs with 
cardinality $|V| = p$, and $E$ is the set of weighted edges. Each 
edge weight in $E$ quantifies the functional connection strength 
between a pair of ROIs such that resulting network $G$ is weighted 
and undirected. For each ROI indexed with $i \in \{1, 2, \ldots, 
p\}$, let $\bmX_i = (X_{1i}, \ldots, X_{ni})^\top \in \mathbb{R}^n$ 
denote the mean BOLD time series averaged over all voxels within that 
region across $n$ time points. Stacking the ROI time series across 
all regions yields the data matrix $\bmX \in \mathbb{R}^{n \times 
p}$, where each row corresponds to the brain activity measured across 
all ROIs at a given time point.

A GGM assumes that the row observations in $\bmX$ are independent and 
identically distributed realizations from a multivariate normal 
distribution $\mathcal{N}_p(\bmmu, \bmSigma)$, with precision matrix 
$\bmOmega = (\omega_{ij})_{p \times p}= \bmSigma^{-1}$. In this 
framework, the network structure is determined by the pairwise Markov 
property; that is, there is no edge between nodes $i$ and $j$ if and 
only if the corresponding variables are conditionally independent 
given all remaining variables,
\[
(i, j) \notin E \Longleftrightarrow \bmX_i \perp \bmX_j \mid 
\bmX_{V \setminus \{i, j\}},
\]
which is equivalent to $\omega_{ij} = 0$. Consequently, the precision 
matrix encodes direct conditional dependencies, allowing GGMs to 
characterize functional connectivity by removing indirect 
associations from other regions. The strength of these direct 
relationships is conventionally summarized by the partial correlation,
\[
\rho_{ij} = -\frac{\omega_{ij}}{\sqrt{\omega_{ii}\omega_{jj}}},
\]
which quantifies the linear association between regions $i$ and $j$ 
after adjusting for all other regions in the brain. This relationship 
highlights the central role of the precision matrix in capturing 
functional interactions and motivates its determination as the 
primary goal in functional brain network analysis.

Let $\bmS$ denote the sample variance and covariance matrix computed 
from $\bmX$. Under the GGM assumption, the log-likelihood function 
for $\bmOmega$ is
\[
\ell(\bmOmega \mid \bmX) = \frac{n}{2}
\left[ \log\det(\bmOmega) - \tr(\bmS\bmOmega) - p\log(2\pi) 
\right],
\]
where $\det(\cdot)$ and $\tr(\cdot)$ denote the determinant and trace 
operators, respectively. When $\bmS$ is positive definite, the 
maximum likelihood estimator (MLE) is given by $\bmOmega = 
\bmS^{-1}$. However, neuroimaging studies typically fall in a 
high-dimensional regime where the number of regions exceeds the 
number of time points ($p > n$), rendering $\bmS$ singular and the 
MLE undefined or numerically infeasible~\citep{sanchez2021combining}. 
To address this challenge and to improve interpretability, 
a variety of estimation methods that impose structural constraints, 
most commonly sparsity, on $\bmOmega$ have been proposed. Such sparse 
solutions not only yield stable estimates in high-dimensional 
settings but also align with the biological expectation that 
functional brain networks are sparsely 
connected~\citep{eavani2015identifying, li2017large}.

\section{Precision Matrix Estimation}
\label{sec:method}

In this section, we review eight well established methods for 
precision matrix estimation. While most of these methods were 
initially developed for regression analysis, they have been adapted 
for use in graphical models. Some methods have been widely applied in 
functional brain network studies, while others have not been 
extensively evaluated in this context. Our review follows a 
systematic format. For each method, we provide its analytical 
expression, a concise explanation of the underlying principle, a 
summary of the recent literature related to neuroimaging 
applications in AD research, and an investigation of available 
\proglang{R} packages.

\subsection{Graphical Lasso (Glasso)}
\label{sec:glasso}

Lasso~\citep{tibshirani1996regression} is a penalized 
regression analysis method for variable selection via $L_1$ 
regularization. In regression analysis, lasso promotes sparsity by
shrinking insignificant regression coefficients 
to zero, which assimilates the sparse structure often observed
in functional brain networks~\citep{eavani2015identifying, 
xue2020estimating}. This similarity has motivated the development of
graphical lasso (or glasso) to estimate precision matrices for 
GGM~\citep{friedman2008sparse, cai2016estimating}. In neuroimaging 
studies, glasso, also known as sparse inverse covariance 
estimation (SICE), is the most widely used method for functional 
connectivity analysis in AD~\citep[e.g.,][]{huang2010learning, 
kim2015testing, ortiz2015exploratory, munilla2017construction, 
kundu2019anovel}.

Glasso estimates $\bmOmega$ by solving
an $L_1$-regularized Gaussian log-likelihood optimization
problem~\citep{friedman2008sparse} defined as
\begin{equation}
	\label{eq:lasso}
	\hatbmOmega_\text{glasso} = \argmin_{\bmOmega} \left\{ - 
	\log\det(\bmOmega)
	+ \tr(\bmS\bmOmega) + \lambda \Vert\bmOmega\Vert_1 \right\},
\end{equation}
where $\lambda \ge 0$ is a tuning parameter 
controlling penalty strength, and
$\Vert\bmOmega\Vert_1 = \sum_{i, j} \vert\omega_{ij}\vert$ is the 
$L_1$ norm of $\bmOmega$. The performance of
$\hatbmOmega_\text{glasso}$ relies on the optimal choice
of~$\lambda$. While the Bayesian information criterion (BIC) is a
common method for selecting~$\lambda$~\citep{yuan2007model,
  friedman2008sparse, gao2012tuning}, other
methods include stability selection~\citep{meinshausen2010stability}, 
extended BIC~\citep{foygel2010extended}, edge false 
discovery rate~\citep{li2013bootstrap}, and 
cross-validation~\citep[CV,][]{vujacic2015acomputationally}. Tuning 
parameter selection has become an important research focus itself, 
though it is not the major concern of this work. To ensure fair
comparison across methods, a five-fold CV, which is applicable to all
considered methods, is used for tuning parameter selection in 
simulations and real data applications for this study.

Several algorithms have been developed to solve the glasso 
optimization 
problem in Equation~\eqref{eq:lasso}. Most notably are the blockwise
coordinate descent~\citep{yuan2007model, banerjee2008model,
  friedman2008sparse} and the alternating direction methods of
multipliers~\citep{boyd2011distributed}. These developments have led
to multiple well-maintained \proglang{R}
packages on the Comprehensive \proglang{R} Archive Network (CRAN): 
\pkg{ADMMsigma}~\citep{pkg:ADMMsigma},
\pkg{CovTools}~\citep{pkg:CovTools}, 
\pkg{glasso}~\citep{pkg:glasso},
\pkg{glassoFast}~\citep{pkg:glassoFast} and 
\pkg{huge}~\citep{pkg:huge}. Additional packages such as 
\pkg{CVglasso}~\citep{pkg:CVglasso} and 
\pkg{GGMncv}~\citep{pkg:GGMncv} also provide implementations of the 
glasso. However, the core estimation routines in these packages 
primarily rely on \pkg{glasso}~\citep{pkg:glasso} and 
\pkg{glassoFast}~\citep{pkg:glassoFast}, 
respectively. Therefore, they were not treated as distinct methods 
in the empirical comparisons of this study.

All six CRAN implementations of glasso produced numerically
indistinguishable precision matrix estimates; computational
efficiency was the only substantive difference. In particular,
\pkg{glassoFast} employs optimized \proglang{FORTRAN} subroutines
extending the blockwise coordinate descent
algorithm~\citep{friedman2007pathwise, sustik2012glassofast} and was
consistently faster than the other packages. Therefore, we use
\pkg{glassoFast} to represent glasso in the simulation studies.

\subsection{Glasso with Ridge Penalty}
\label{sec:ridge}

A different regularization is ridge penalty~\citep{hoerl1970ridge}, 
which is also known as $L_2$ regularization or Tikhonov 
regularization. One main difference between ridge regularization and 
$L_1$ regularization is that ridge penalty does not penalize ``weak 
connectivity'' (e.g., low correlations between ROIs) to zero. 
Accordingly, ridge penalty appears more suitable for non-sparse 
networks, such as brain networks generated from positron emission 
tomography (PET) imaging~\citep{veronese2019covariance, 
huang2020anovel}. Recently, \citet{liegeois2020revisiting} used 
$L_2$-regularized precision matrix to estimate functional 
connectivity of (relatively denser) sub-networks (e.g., 
frontoparietal network and somatomotor network), and discovered that 
$L_2$ regularization would outperform competing methods for 
precision matrix estimation when very few data points were available 
for the analysis. This finding is consistent with the conclusion of 
an independent study~\citep{pervaiz2020optimising}, where the 
authors pointed out that $L_2$ regularization was preferred in 
functional connectivity estimation when scanning session duration 
was short (i.e., small sample size) since $L_2$ regularization was 
more efficient and stable.

The precision matrix estimator based on ridge penalty is to solve 
the following optimization problem:
\begin{equation}
	\label{eq:ridge}
	\hatbmOmega_\text{ridge} = \argmin_{\bmOmega} \left\{
	- \log\det(\bmOmega) + \tr(\bmS\bmOmega)
	+ \frac{1}{2}\lambda\Vert\bmOmega\Vert_2^2 \right\}
\end{equation}
where  $\lambda \ge 0$ is a tuning parameter, and 
$\Vert\bmOmega\Vert_2 = (\sum_{i, j} \omega_{ij}^2)^{1/2}$ is the 
$L_2$ norm of $\bmOmega$. This optimization problem~\eqref{eq:ridge} 
has a closed-form solution~\citep{vanwieringen2016ridge}:
\[
\hat{\bmOmega}_\text{ridge} = \left\{\left[\lambda\bmI + 
\frac{1}{4}\bmS^2 \right]^{1/2} + \frac{1}{2}\bmS \right\}^{-1},
\]
where $\bmI$ is an identity matrix of dimension $p$ by $p$.

Four \proglang{R} packages, 
\pkg{ADMMsigma}~\citep{pkg:ADMMsigma}, 
\pkg{GLassoElnetFast}~\citep[a GitHub package not yet available on 
the CRAN,][]{kovacs2021graphical}, 
\pkg{porridge}~\citep{pkg:porridge} and 
\pkg{rags2ridges}~\citep{pkg:rags2ridges}, 
provide functions for $L_2$-regularized optimization. 
Our experiment showed that the resulting precision matrices were 
identical, but \pkg{rags2ridges} required substantially less
computing time. Therefore, we use \pkg{rags2ridges} for 
$L_2$ regularization in the simulation studies.

\subsection{Graphical Elastic Net}
\label{sec:elnet}

Elastic net~\citep{zou2005regularization} is a trade-off approach 
integrating the advantages of $L_1$ and $L_2$ regularization to 
achieve both bias and variance reduction. Graphical elastic net aims 
to solve an optimization problem based on a penalized Gaussian 
log-likelihood containing both $L_1$ and $L_2$ regularized terms 
such that the estimated precision matrix is expected to balance 
network sparsity and desired graphical model structure. Albeit its
flexibility, there is limited research in the literature utilizing 
graphical elastic net to construct functional brain networks. 
\citet{ryali2012estimation} applied graphical elastic net to 
functional connectivity estimation and reported that the combination 
of $L_1$ regularization (facilitating network sparsity) and $L_2$ 
regularization (promoting method sensitivity) would improve model 
interpretability. Later, \citet{guo2018resting} used 
graphical elastic net to build hyper-networks for the brain, and 
found it outperformed glasso in classification analyses.

Graphical elastic net estimates $\bmOmega$ by solving the following 
optimization problem~\citep{kovacs2021graphical} expressed as
\begin{equation*}
	\label{eq:elnet}
	\hatbmOmega_\text{elnet} = \argmin_{\bmOmega} 
	\left\{-\log\det(\bmOmega)
	+ \tr(\bmS\bmOmega) + \lambda \left[
	\gamma\Vert\bmOmega\Vert_1 + 
	\frac{1-\gamma}{2}\Vert\bmOmega\Vert_2^2
	\right] \right\},
\end{equation*}
where $\gamma \in [0,1]$ is a secondary tuning parameter to 
balance $L_1$ and $L_2$ penalties.

Two \proglang{R} packages, 
\pkg{ADMMsigma}~\citep{pkg:ADMMsigma} and
\pkg{GLassoElnetFast}~\citep{kovacs2021graphical}, implement
graphical elastic net. Through our experiments, 
\pkg{ADMMsigma} utilizing the alternating direction method of 
multipliers~\citep{boyd2011distributed} demonstrated 
better performance under the evaluation metrics considered in our 
study, and required less computational time in high-dimensional 
settings. Therefore, only results from \pkg{ADMMsigma} are reported
in the simulation studies.

\subsection{Adaptive Graphical Lasso}
\label{sec:alasso}

It is arguable that lasso-based approaches may suffer from the 
limitation of uniformly applying $L_1$ regularization to all 
objectives, tentatively resulting in 
overfitting~\citep{fan2001variable}. This limitation motivates the 
development of adaptive lasso~\citep{zou2006adaptive} which 
facilitates uneven weights to regularized terms. In an unpublished 
preprint, \citet{zhou2009adaptive} proposed a method (i.e., adaptive 
graphical lasso) that leveraged the concept of adaptive lasso to 
estimate the precision matrix of GGM. However, to the best of our 
knowledge, the application of adaptive graphical lasso in 
constructing functional brain networks is extremely limited in the 
literature. Until recently, \citet{wodeyar2022structural} used a 
simplified adaptive graphical lasso algorithm to build brain 
connectomes based on magnetoencephalography (MEG) data.

The precision matrix estimator based on adaptive graphical
lasso~\citep{fan2009network} is the solution to the following 
optimization problem:
\begin{equation*}
	\label{eq:adaptive}
	\hatbmOmega_\text{adapt} = \argmin_{\bmOmega} \left\{
	- \log\det(\bmOmega) + \tr(\bmS\bmOmega)
	+ \lambda \Vert \bmV \otimes \bmOmega \Vert_1 \right\},
\end{equation*}
where $\otimes$ denotes element-wise multiplication for matrices, 
and $\bmV := (v_{ij})_{p \times p}$ is a matrix of adaptive weights. 
There are two conventional approaches for selecting adaptive 
weights: (1) $v_{ij} = 1/|s_{ij}|$, where $s_{ij}$ is the 
corresponding entry in the sample variance and covariance matrix 
$\bmS$~\citep{bien2011sparse}, and (2) $v_{ij} = 
1/|\tilde{\omega}_{ij}|^{\gamma}$, where $\tilde{\omega}_{ij}$ 
is the corresponding entry in a consistent precision matrix 
estimator $\tildebmOmega$ (e.g., 
$\hat{\bm{\Omega}}_{\text{glasso}}$) 
and $\gamma > 0$ is an independent tuning parameter. In this 
research, we fixed $\gamma = 0.5$ (for adaptive graphical lasso) as 
suggested by~\citet{fan2009network}.

Three \proglang{R} packages, \pkg{glasso}~\citep{pkg:glasso}, 
\pkg{glassoFast}~\citep{pkg:glassoFast} and 
\pkg{GLassoELnetFast}~\citep{pkg:GLassoElnetFast}, can be used to 
implement adaptive graphical lasso after minor modifications of the 
existing functions therein. Our numeric experiments revealed that 
these three packages produced similar precision matrix estimates, 
but the estimation accuracy varied with respect to the selection of 
weighting matrix $\bm{V}$ (and also $\tildebmOmega$). More 
specifically, we experimented $v_{ij} = 1 / \vert 
s_{ij} \vert$, and $v_{ij} = 1 / \vert \tilde{\omega}_{ij} 
\vert^{0.5}$ with $\tildebmOmega$ chosen from $\bmS^{-1}$ (where 
applicable), $\hatbmOmega_\text{glasso}$, $\bmS_\text{lin}^{-1}$ 
(the inverse of the Ledoit--Wolf linear shrinkage 
estimator~\citep{ledoit2004well} of the covariance matrix $\bmSigma$) 
and $\bmS_\text{nl}^{-1}$ (the inverse of the Ledoit--Wolf non-linear 
shrinkage estimator~\citep{ledoit2015spectrum, ledoit2017numerical} 
of covariance matrix $\bmSigma$), respectively. Our findings 
suggested that $\tildebmOmega = \bmS^{-1}$ performed best in 
low-dimensional settings, while $\tildebmOmega= \bmS^{-1}_\text{lin}$ 
exhibited more stability in high-dimensional settings. In addition, 
our experiments showed that the \pkg{glassoFast} implementation is 
more computationally efficient. Therefore, we report subsequent 
simulation results for low- and high-dimensional settings using these 
respective estimator-implementation combinations.

Finally, we note that the \proglang{R} package 
\pkg{GGMncv}~\citep{pkg:GGMncv} also provides adaptive graphical 
lasso implementation, but it is built directly on 
\pkg{glassoFast}~\citep{pkg:glassoFast}. Therefore, we do not 
consider it separately in our evaluation.

\subsection{SCAD}
\label{sec:scad}

The SCAD (\underline{s}moothly \underline{c}lipped 
\underline{a}bsolute \underline{d}eviation) 
method~\citep{fan2001variable} is an alternative that overcomes the 
limitation of even penalization for glasso. Specifically, SCAD 
introduces a non-convex penalty based on differentiated weights 
given by
\begin{equation}
	\label{eq:scad_weight}
	\psi^{\prime}_{\lambda, \gamma}(x) = \lambda \bm{1}(|x| \leq 
	\lambda) + \frac{(\gamma\lambda - |x|)}{\gamma - 1} 
	\bm{1}(\lambda < |x| < \gamma\lambda),
\end{equation}
where $\bm{1}(\cdot)$ is the standard indicator function, the tuning 
parameter $\lambda$ is defined identically to that for glasso, and 
the other tuning parameter $\gamma > 2$ governs the concavity of 
penalization. The first term on the right-hand-side of 
Equation~\eqref{eq:scad_weight} warrants a standard penalty to small 
elements, whereas the second term applies a reduced penalty to 
moderate elements. The tuning parameter $\gamma$ identifies which 
elements will receive reduced penalty and subsequently which 
(larger) elements will not be penalized. One can show that the SCAD 
estimator is asymptotically equivalent to the glasso estimator as 
$\gamma \to \infty$. In practice, a recommended value for~$\gamma$ is 
$3.7$~\citep{fan2001variable}, which will be used in our simulations. 
Although SCAD has not yet been popularized in functional connectivity 
analysis, \citet{zhu2018sparse}, in an independent review article, 
posited the potential of SCAD to reduce false positive connections 
when applied to functional brain network research. More recently, 
\citet{hrybouski2023aging} used SCAD to estimate the network 
structure of medial temporal lobe in an AD 
study.

The precision matrix estimator based on the SCAD 
method~\citep{fan2009network} can be obtained by solving the 
following optimization problem:
\begin{equation*}
	\label{eq:scad}
	\hatbmOmega_\text{scad} = \argmin_{\bmOmega} \left\{
	- \log\det(\bmOmega) + \tr(\bmS\bmOmega) + \sum_{i, j} 
	\psi_{\lambda, \gamma}(|\omega_{ij}|) \right\}.
\end{equation*}
The optimization of $\hat{\bm\Omega}_{\text{scad}}$ is performed 
iteratively using the local linear approximation (LLA) strategy 
introduced by~\citet{fan2009network}. Under this framework, the 
nonconvex SCAD penalty is approximated by a sequence of weighted 
lasso problems, where the weights are determined by the derivative 
of the penalty evaluated at the current estimate. This formulation 
allows the use of standard glasso algorithms by incorporating the 
weight matrix as element-wise tuning parameters, while the core 
optimization routine remains unchanged. Consequently, several 
existing \proglang{R} packages developed for glasso can be adapted to 
implement SCAD.

Implementation of SCAD also requires an initial estimate of the 
precision matrix. We evaluated multiple initialization strategies and 
found that using the glasso estimator, 
$\hat{\bm\Omega}_{\text{glasso}}$, as the starting estimate 
consistently yielded the most accurate results. Among available 
implementations, the \proglang{R} package \pkg{glassoFast} provided 
the best computational efficiency. Accordingly, this implementation 
was adopted to obtain the SCAD results to be reported in our 
simulation studies.

\subsection{MCP}
\label{sec:mcp}

Similar to the conceptual mechanism of SCAD, MCP 
(\underline{m}inimax \underline{c}oncave \underline{p}enalty) 
proposed by~\citet{zhang2010nearly} is an alternative incorporating 
a non-convex penalty and assigning unequal penalties to targets. To 
distinguish, we introduce a different notation defining 
differentiated weights for MCP as 
follows:
\begin{equation}
	\label{eq:mcp}
	\xi^{\prime}_{\lambda, \gamma}(x) = \left(\lambda - 
	\frac{|x|}{\gamma}\right)
	\bm{1}(|x| \leq \gamma\lambda)
\end{equation}
with parameters $\lambda > 0$ and $\gamma > 1$ respectively controlling 
sparsity and concavity. The interpretation of $\gamma$ is analogous 
to that in Equation~\eqref{eq:scad_weight}, and a conventional 
selection of $\gamma$ for MCP is $3$, as suggested by 
\citet{breheny2011coordinate}. Equation~\eqref{eq:mcp} implies that 
MCP reduces penalization rate with the increase of $|x|$. A 
significant distinction (of MCP) from SCAD is that the change of 
penalization rate for MCP takes place continuously, but for SCAD, the 
change does not occur until $|x|$ reaches a threshold (i.e., 
$\lambda$).

The estimator for precision matrix based on MCP is formulated as
\begin{equation*}
	\hatbmOmega_\text{mcp} = \argmin_{\bmOmega} \left\{
	- \log\det(\bmOmega) + \tr(\bmS\bmOmega) + \sum_{i, j} 
	\xi_{\lambda,\gamma}(|\omega_{ij}|) \right\}.
\end{equation*}

To date, we have not yet noticed any existing research utilizing MCP for 
brain network connectivity analysis in AD, although it has 
been applied in GGM-based models for gene co-expression networks 
in breast cancer studies~\citep{wang2016precision}. It is therefore of
value to examine its performance and adaptability for  
graphical modeling of functional brain connectome.

Given its structural similarity to SCAD, MCP estimation can be 
implemented using existing \proglang{R} packages developed for 
glasso-based optimization. In our simulation studies, MCP was 
implemented using the \pkg{glassoFast} package, with the glasso 
estimate $\hat{\bm{\Omega}}_{\text{glasso}}$ used as the initial 
value for the iterative MCP optimization.

\subsection{CLIME}
\label{sec:clime}

All the methods we have reviewed so far are essentially based on 
penalized Gaussian log-likelihood optimization, accommodated by 
different types of regularization terms. Another class of methods 
based on constrained $\ell_1$ minimization have become a competitive 
alternative evidenced by the fact that constrained $\ell_1$ 
minimization is effective for constructing sparse 
signals~\citep{candes2007dantzig}. As shown 
by~\citet{cai2011aconstrained}, methods based on constrained 
$\ell_1$ minimization are particularly attractive because the 
associated convex optimization problems can be solved column by 
column (i.e., linear programming), leading to a significant reduction 
of computation time. One typical approach from this class 
specifically for precision matrix estimation is CLIME 
(\underline{c}onstrained 
\underline{$\ell_1$}-minimization for \underline{i}nverse 
\underline{m}atrix \underline{e}stimation), firstly introduced 
by~\citet{cai2011aconstrained}. Since recently, CLIME has gradually 
become more frequently used for precision matrix estimation in 
GGM-based functional connectivity research for
AD~\citep{wang2016anefficient, chen2022simultaneous}, but its 
performance in functional brain network research compared to 
glasso-based methods remains unclear, which partly motivates the 
present research.

The precision matrix estimator for CLIME is the solution to the 
following convex optimization problem:
\begin{equation}
	\label{eq:clime}
	\hatbmOmega_\text{clime} = \argmin \Vert\bmOmega\Vert_1 \qquad
	\text{subject to } \Vert\bmS\bmOmega-\bmI\Vert_\infty \leq 
	\lambda,
\end{equation}
where $\Vert\bmS\bmOmega - \bmI\Vert_\infty$ returns the maximum 
absolute value of the elements in $\bmS\bmOmega - \bmI$, and the 
tuning parameter $\lambda$ is interpreted similarly to that for 
glasso. It is worth mentioning that the estimator 
$\hatbmOmega_\text{clime}$ is not necessarily symmetric. For 
undirected functional brain network research that requires a 
symmetric precision matrix, one suggests setting $\omega_{ij} = 
\omega_{ji} = \omega_{ij} \bm{1}(|\omega_{ij}| \leq |\omega_{ji}|)
+ \omega_{ji} \bm{1}(|\omega_{ij}| > |\omega_{ji}|)$ for each pair 
of $i$ and $j$ in $\hatbmOmega_\text{clime}$.

Two \proglang{R} packages, 
\pkg{clime}~\citep{pkg:clime} and \pkg{flare}~\citep{pkg:flare}, 
provide functions implementing CLIME. Our experiments 
showed that the results from \pkg{flare} were closer to the ground 
truth when evaluated by the Frobenius norm.
Besides, the Dantzig selector used in \pkg{flare} 
promotes sparser solutions than alternative algorithms for solving the 
optimization problem in Equation~\eqref{eq:clime}. Hence, only
results from \pkg{flare} are reported in the subsequent 
simulation studies.

\subsection{TIGER}
\label{sec:tiger}

TIGER (\underline{T}uning-\underline{I}nsensitive \underline{G}raph 
\underline{E}stimation and \underline{R}egression), proposed by 
\citet{liu2017tiger}, is a robust precision matrix estimation method 
designed to reduce sensitivity to tuning parameter selection. Unlike 
most glasso-based approaches, TIGER is built upon the 
square-root lasso framework~\citep{belloni2011square}. Specifically, 
the coefficient matrix $\bmB \in \mathbb{R}^{p \times p}$ is obtained 
by solving
\[
\bmB = \arg\min_{\bmB} \|\bmX - \bmX\bmB\|_{2,1} + \lambda \|\bmB\|_1
\]
subject to $B_{jj} = 0$ for $j = 1, 2, \ldots, p$, where 
$\bmX \in \mathbb{R}^{n \times p}$ is the data matrix and
\[
\|\bmX - \bmX\bmB\|_{2,1} 
= \sum_{j = 1}^p 
\left( \sum_{k = 1}^{n} 
\{(\bmX - \bmX\bmB)_{kj}\}^2 \right)^{1/2}
\]
denotes the $\ell_{2,1}$ norm of the residual matrix with $\lambda = 
\sqrt{\log{p} / n}$.

Let $\sigma_j^2$ denote the residual variance from regressing the 
$j$-th column of $\bmX$ (i.e., $\bmX_j$) on the remaining columns 
$\bmX_{-j} \in \mathbb{R}^{n \times (p-1)}$. The precision matrix 
estimator~\citep{liu2017tiger} is then constructed as
\[
(\hat{\bmOmega}_{\text{tiger}})_{jj} = \sigma_j^{-2}, 
\qquad \text{and} \qquad
(\hat{\bmOmega}_{\text{tiger}})_{ij} 
= -\sigma_j^{-2} B_{ij}
\]
for $i \neq j$. Similar to CLIME, the TIGER estimator is constructed 
column-wise and does not ensure a symmetric matrix output. 
Consequently, a post hoc symmetrization step is required for 
undirected network analysis, typically using the symmetrization 
procedure developed for CLIME.

In neuroimaging applications, TIGER has been used to estimate
functional connectivity networks from high-dimensional fMRI
data~\citep{pircalabelu2015afocused} and to characterize
disease-related
network structure in AD studies~\citep{dyrba2018comparison}. 
Two \proglang{R} packages, \pkg{flare}~\citep{pkg:flare} and 
\pkg{huge}~\citep{pkg:huge}, provide implementations of TIGER. In 
our additional experiments not presented in the paper, \pkg{flare} 
yielded smaller Frobenius norm values, indicating better element-wise 
estimation accuracy, whereas \pkg{huge} gave lower Kullback--Leibler 
divergence values. These differences may be attributed to variations 
in algorithmic implementation and tuning parameter selection 
strategies. Since our primary focus is on 
precision matrix estimation accuracy,  we used \pkg{flare} for TIGER
in the simulation study.

For fair cross-method comparison, we adopt a unified tuning strategy 
for all estimators, including TIGER. The theoretically motivated 
choice $\lambda = \sqrt{\log(p)/n}$ corresponds to the asymptotic rate 
derived under high-dimensional conditions \citep{liu2017tiger}, which 
guarantees desirable theoretical properties. However, asymptotic 
prescriptions do not necessarily yield optimal finite-sample 
performance in practice. Although TIGER is designed to exhibit reduced 
sensitivity to the tuning parameter, we selected $\lambda$ via 
five-fold cross-validation, consistent with the procedure applied to 
the other methods. This intentional choice ensures comparability across 
estimators and evaluates each method under an empirically optimized 
configuration within the same tuning framework.

\subsection{Software Summary}
\label{sec:pkg}

\begin{sidewaystable}[tbp]
	\centering
	\caption{Summary of precision matrix estimation methods, 
	corresponding \proglang{R} packages, and tuning parameter 
	selection strategies.}
	\footnotesize
	\label{tab:method_summary}
	
	\begin{tabular*}{\textheight}{@{\extracolsep{\fill}}p{4.0cm}p{2.8cm}p{3.4cm}p{3.7cm}p{6cm}}
		\toprule
		\textbf{Method (Full Name)} & \textbf{Acronym} & 
		\textbf{Reference(s)} & \textbf{R Packages} & \textbf{Tuning 
		Selection} \\
		\midrule
		
		Graphical Lasso 
		& Glasso 
		& \citet{friedman2008sparse} 
		& \pkg{ADMMsigma}; \pkg{CovTools}; \pkg{glasso}; 
		\pkg{glassoFast}; \pkg{huge}; \pkg{CVglasso}; \pkg{GGMncv}
		& BIC; $10$-fold CV (Neg LogLik, SPE) \\
		
		Glasso with Ridge 
		& Ridge 
		& \citet{vanwieringen2016ridge} 
		& \pkg{ADMMsigma}; \pkg{GLassoElnetFast}; \pkg{porridge}; 
		\pkg{rags2ridges} 
		& LOOCV (Neg LogLik) \\
		
		Graphical Elastic Net 
		& Elnet 
		& \citet{kovacs2021graphical} 
		& \pkg{ADMMsigma}; \pkg{GLassoElnetFast}
		& $5$- or $10$-fold CV (Neg LogLik, MSPE) \\
		
		Adaptive Graphical Lasso 
		& Glasso (Adapt)
		& \citet{fan2009network} 
		& \pkg{glasso}; \pkg{glassoFast}; \pkg{GLassoElnetFast} 
		& $5$- or $6$-fold CV (RPE; specialized loss) \\
		
		Smoothly Clipped Absolute Deviation
		& SCAD 
		& \citet{fan2001variable} 
		& \pkg{glasso}; \pkg{glassoFast}; \pkg{GLassoElnetFast}
		& $5$- or $6$-fold CV (MPSE; specialized loss); GCV \\
		
		Minimax Concave Penalty 
		& MCP 
		& \citet{zhang2010nearly} 
		& \pkg{glasso}; \pkg{glassoFast}; \pkg{GLassoElnetFast}
		& $10$-fold CV (MPSE) \\
		
		Constrained $\ell_1$ Minimization for Inverse Matrix 
		Estimation
		& CLIME 
		& \citet{cai2011aconstrained} 
		& \pkg{clime}; \pkg{flare} 
		& $5$-fold CV (Neg LogLik; $L_2$ trace) \\
		
		Tuning-Insensitive Graph Estimation and Regression
		& TIGER 
		& \citet{liu2017tiger} 
		& \pkg{flare}; \pkg{huge} 
		& Stability selection; $5$-fold CV (Neg LogLik; $L_2$ trace) 
		\\
		
		\bottomrule
	\end{tabular*}
	
	\vspace{2mm}
	\begin{minipage}{\textheight}
		\footnotesize
		BIC: Bayesian information criterion; CV: cross-validation; 
		Neg LogLik: negative log-likelihood; SPE: squared prediction 
		error;  LOOCV: Leave-one-out CV; MSPE: mean squared 
		prediction error; RPE: relative prediction error; 
		GCV: generalized cross-validation.
	\end{minipage}
	
\end{sidewaystable}

Table~\ref{tab:method_summary} summarizes the available \proglang{R}
implementations for the precision matrix estimation methods reviewed
in this section at the time of writing. Most 
implementations are distributed through CRAN, ensuring standardized 
installation procedures and version control, whereas a smaller number 
are maintained on GitHub or other repositories. The software 
ecosystem is heterogeneous, with several methods supported by multiple 
packages that differ in algorithmic implementation and computational 
design. While some packages focus on a single estimator, others 
provide unified interfaces for multiple penalties within the GGM 
framework. This diversity reflects the active development of 
high-dimensional graphical modeling tools and underscores the 
importance of transparent software documentation and reproducibility 
in functional connectivity research.

\section{Simulation Studies}
\label{sec:sim}

Although the aforementioned estimation methods have been supported by 
extensive theoretical developments and demonstrated in a variety of 
applications, systematic comparisons of their performance for 
functional connectivity estimation remain limited. In this section, 
we conduct comprehensive data-driven simulations to evaluate the 
methods introduced in Section~\ref{sec:method}. The performance of 
each method is assessed from multiple perspectives, including 
estimation accuracy, structural recovery, and computational 
efficiency.

\subsection{Synthetic Data Generation and Simulation Setup}
\label{sec:data_gen}

A data-driven strategy was adopted to construct the true precision 
matrix, $\bmOmega$. The sample size was fixed at $n = 180$, while 
multiple network dimensions corresponding to different brain 
parcellations were considered, with $p \in \{100, 200, 400\}$. For 
each value of $p$, we computed the sample covariance matrix from the 
real, preprocessed fMRI data and applied the glasso to 
obtain an estimated precision matrix, which was treated as the 
ground-truth matrix~$\bmOmega$. Synthetic datasets $\bmX_i$ were 
then generated from a multivariate normal distribution with mean 
$\bm{0}$ and covariance matrix $\bmOmega^{-1}$. The number of 
simulation runs was set to $100$.

As discussed in Section~\ref{sec:method}, tuning parameter selection 
for the regularization parameter~$\lambda$ differs across methods and 
may also vary by algorithm implementation. 
Table~\ref{tab:method_summary} 
summarizes tuning strategies recommended in the literature, where 
five-fold cross-validation is most commonly used. In our simulations, 
the tuning parameter $\lambda$ was selected via five-fold 
cross-validation based on the log-likelihood loss. For methods that 
involve an additional tuning parameter, denoted by $\gamma$ (with 
notation varying across the literature), its value was specified 
based on commonly adopted recommendations from prior studies: $\gamma 
= 0.5$ for adaptive graphical 
lasso~\citep{fan2009network}, $\gamma = 3.7$ for 
SCAD~\citep{fan2001variable}, and $\gamma = 3$ for 
MCP~\citep{breheny2011coordinate}. For graphical elastic net, a grid 
search was conducted over $\gamma \in \{0.1, 0.2, \ldots, 0.9\}$ in a 
manner analogous to the selection procedure for~$\lambda$.

\subsection{Performance Assessment}
\label{sec:assess}

For performance evaluation, we adopted three commonly used metrics 
following the recommendations of~\citet{yuan2007model, 
fan2009network, cai2011aconstrained}. Specifically, we considered the 
Frobenius norm (F-norm), the Kullback--Leibler (K-L) divergence, and 
the $F_1$ score.

The Frobenius norm is defined as
\[
\Delta_F = \|\bmOmega - \hat{\bmOmega}\|_F = \left( \sum_{i=1}^{p} 
\sum_{j = 1}^{p} (\omega_{ij} - \hat{\omega}_{ij})^2 \right)^{1/2},
\]
which measures the overall element-wise discrepancy between the 
estimated and true precision matrices. Larger values of $\Delta_F$ 
indicate lower estimation accuracy.

The Kullback--Leibler divergence is given by
\[
\Delta_{KL} = \mathrm{tr}(\bmSigma \hat{\bmOmega}) 
- \log \det(\bmSigma \hat{\bmOmega}) - p,
\]
which quantifies the information loss incurred when the estimated 
precision matrix is used in place of the truth. Larger values of 
$\Delta_{KL}$ indicate worse model fit.

Lastly, we computed the $F_1$ score,
\[
F_1 = \frac{2\,\mathrm{TP}}{2\,\mathrm{TP} + \mathrm{FP} + 
	\mathrm{FN}},
\]
where TP, FP, and FN denote the numbers of true positive, false 
positive, and false negative edges in $\hat{\bmOmega}$, respectively. 
The $F_1$ score ranges from $0$ to $1$, with values closer to $1$ 
indicating more accurate recovery of the true network structure. 
Because the $F_1$ score is based on the presence or absence of edges, 
it does not account for the magnitudes of the estimated precision 
matrix entries. However, it provides a useful summary of the ability 
to correctly identify the network sparsity pattern, balancing 
sensitivity and specificity in edge detection.

These metrics assess performance from complementary perspectives, 
including estimation accuracy, information loss, and network 
recovery. In addition, we report the average computational time 
(across $100$ simulation replicates) for each method. All simulations 
were performed on a system equipped with an AMD EPYC 7713 
processor running at 2.6 GHz with 16 MB of cache. Each simulation 
replicate was executed as an independent, single-threaded job on a 
separate computing node.

To facilitate reproducibility, we developed an \proglang{R} package, 
\pkg{spice}~\citep{pkg:spice}, that integrates the implementations of 
all surveyed methods, available at 
\url{https://github.com/Carol-seven/spice}. The online supplement for 
this manuscript provides example simulation scripts used to reproduce 
the results reported in this study.

\subsection{Simulation Results}
\label{sec:sim_res}

\begin{table}[tbp]
	\centering
	\caption{Performance of the surveyed methods with a fixed sample 
	size of $n = 180$ based on $100$ simulation replicates.}
	\label{tab:sim_res}
	\begin{tabular*}{\textwidth}{@{\extracolsep{\fill}}llrrrrr}
		\toprule
		Dim ($p$) & Method & F-norm & K-L div  & $F_1$ 
		score & 
		Sparsity & Time (in sec) \\ 
		\midrule
		100 & Glasso & 10.9320 & 9.8998 & 0.4806 & 0.8073 & 0.8235 \\ 
		& Ridge & 11.4456 & 13.5859 & 0.3149 & 0.0000 & 0.5276 \\ 
		& Elnet & 7.5513 & 6.1929 & 0.4729 & 0.7911 & 97.2185 \\ 
		& Adapt & 8.7536 & 8.7100 & 0.4079 & 0.8748 & 1.2443 \\ 
		& SCAD & 6.6344 & 6.1904 & 0.4574 & 0.8629 & 1.6255 \\ 
		& MCP & 6.5187 & 6.0930 & 0.4554 & 0.8870 & 1.7688 \\ 
		& CLIME & 11.0272 & 31.3495 & 0.4335 & 0.8673 & 255.8024 \\ 
		& TIGER & 6.9780 & 28.9638 & 0.4703 & 0.8051 & 291.5353 \\ 
		\midrule
		200 & Glasso & 14.2233 & 20.0114 & 0.3983 & 0.8707 & 4.5862 \\ 
        & Ridge & 16.9862 & 38.0528 & 0.1981 & 0.0000 & 3.1703 \\ 
        & Elnet & 10.4015 & 14.0222 & 0.3947 & 0.8750 & 804.1154 \\ 
        & Adapt & 12.0804 & 18.7178 & 0.3553 & 0.9286 & 12.6186 \\ 
        & SCAD & 9.2320 & 13.4425 & 0.3886 & 0.9087 & 8.6720 \\ 
        & MCP & 9.0216 & 13.1853 & 0.3929 & 0.9311 & 8.0197 \\ 
        & CLIME & 12.2641 & 66.2096 & 0.3473 & 0.8853 & 2200.8763 \\ 
        & TIGER & 11.4906 & 122.1860 & 0.4018 & 0.8772 & 4356.7173 \\ 
		\midrule
		400 & Glasso & 16.1786 & 37.9359 & 0.3016 & 0.9033 & 24.6609 \\ 
        & Ridge & 22.1975 & 96.4033 & 0.0996 & 0.0000 & 20.5801 \\ 
        & Elnet & 13.7746 & 30.4218 & 0.3595 & 0.9353 & 7121.8989 \\ 
        & Adapt & 13.9957 & 33.7803 & 0.3300 & 0.9553 & 46.9138 \\ 
        & SCAD & 12.0888 & 27.9775 & 0.3458 & 0.9403 & 44.7093 \\ 
        & MCP & 11.6575 & 26.8984 & 0.3631 & 0.9581 & 44.7235 \\ 
        & CLIME & 15.2192 & 153.2777 & 0.2701 & 0.9182 & 22395.7244 \\ 
        & TIGER & 14.5099 & 172.0717 & 0.3014 & 0.8684 & 7188.0763 \\ 
		\bottomrule
	\end{tabular*}
\end{table}

Table~\ref{tab:sim_res} summarizes the simulation results, and 
reveals several consistent patterns across network dimensions. 
Overall, the nonconvex penalization methods, MCP and SCAD, achieved 
the highest estimation accuracy, as indicated by the smallest F-norms 
and K-L divergence values across most settings. MCP consistently 
showed slightly better performance than SCAD, although the 
differences were modest. Elastic net (Elnet) performed competitively, 
with accuracy comparable to SCAD and MCP in low- to 
moderate-dimensional settings, but its performance deteriorated as 
the dimension increased and came at the cost of substantially higher 
computational burden. TIGER showed competitive performance in low 
dimensions and remained comparable to elnet in terms of estimation 
accuracy as the dimension increased. In comparison, the standard 
glasso, adaptive glasso (adapt), and CLIME presented larger 
estimation errors but demonstrated stable performance across 
dimensions. Among these methods, the adaptive 
glasso generally provided the best accuracy. Ridge consistently 
showed the poorest estimation performance across all settings.

From a structural recovery perspective, sparsity-inducing methods 
(glasso, adaptive glasso, SCAD, MCP, and elastic net) achieved 
comparable $F_1$ scores, with MCP typically attaining the highest 
values and producing the sparsest solutions. CLIME and TIGER achieved 
reasonable edge recovery in some settings. However, their 
K-L divergence values, particularly for TIGER, were 
substantially larger, indicating considerable information loss 
despite acceptable structural recovery. In contrast, ridge estimation 
produced dense solutions (zero sparsity) and relatively low $F_1$ 
scores, reflecting its inability to recover sparse network structures.

In terms of computational efficiency, ridge was the fastest method 
across all dimensions. The adaptive glasso, SCAD, and MCP 
were also computationally efficient and showed good scalability as 
the dimension increased, with adaptive GLASSO being faster in low 
dimensions and SCAD and MCP becoming more efficient in higher 
dimensions. The standard glasso required moderate 
computation time. In contrast, elastic net, CLIME, and TIGER were 
substantially more computationally intensive, and their computational 
cost increased rapidly with network dimension. Overall, these results 
suggest that MCP and SCAD provide the best statistical accuracy, with 
MCP showing a slight advantage over SCAD. Compared with the standard 
glasso, the adaptive glasso offers improved accuracy together with 
favorable computational efficiency. Elastic net remains competitive 
in moderate-dimensional settings, but is less attractive in practice 
due to its relatively higher computational cost.

\begin{figure}[tbp]
\centering
\label{fig:boxplot}
\includegraphics[width=\textwidth]{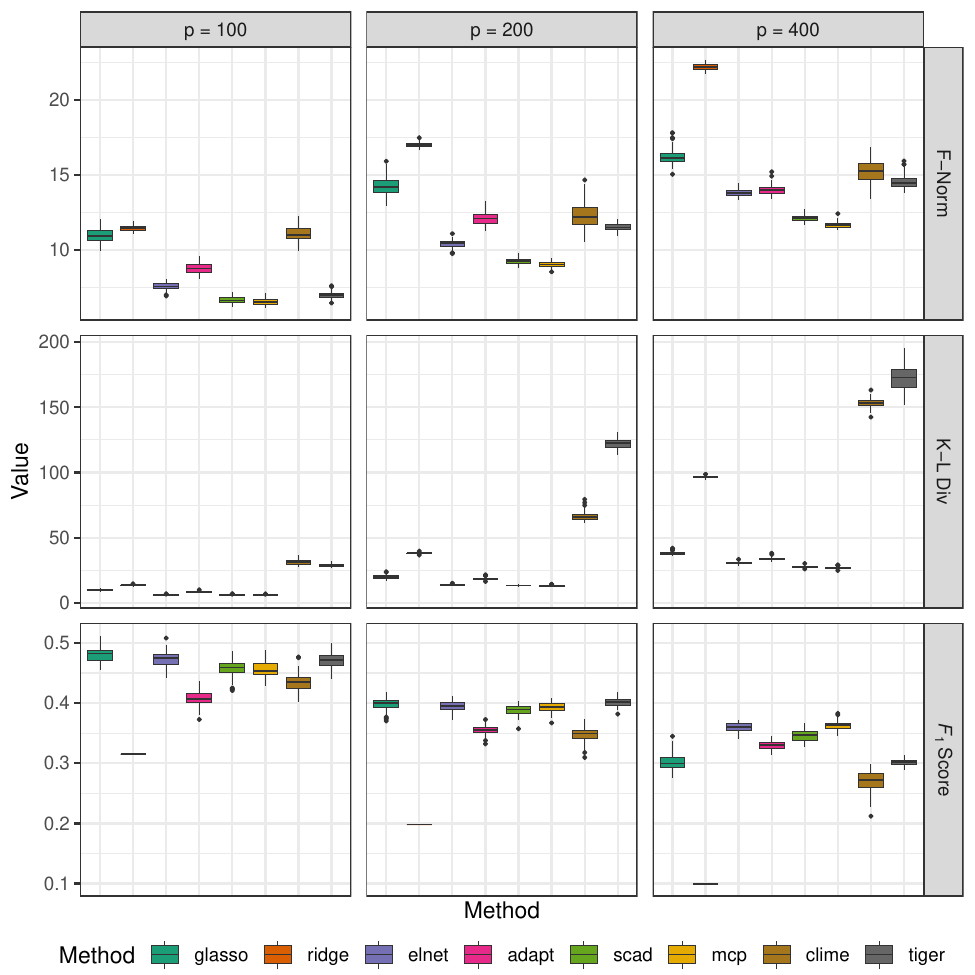}
\caption{Boxplots of the estimation performance metrics (Frobenius 
norm, Kullback--Leibler divergence, and $F_1$ score) for all methods 
across different network dimensions (with fixed $n = 180$), based on 
$100$ simulation replicates.}
\end{figure}

The boxplots in Figure~\ref{fig:boxplot} further illustrate the 
distributional characteristics of the performance metrics across 
simulation replicates and provide additional insight beyond the 
average results reported in Table~\ref{tab:sim_res}. Overall, MCP and 
SCAD not only achieved lower median F-norm and K-L divergence 
values but also exhibited relatively small interquartile ranges, 
indicating stable performance across simulations. Although Elnet 
showed comparable median accuracy in low to moderate dimensions, its 
variability increased as the dimension grew. In contrast, the 
standard GLASSO and adaptive GLASSO displayed moderate dispersion but 
remained stable across different settings. The variability patterns 
are more significant for CLIME and TIGER, whose K-L divergence values 
exhibit substantial spread, particularly in higher dimensions, 
reinforcing instability in estimation despite reasonable edge 
recovery.

Similar trends are observed in the $F_1$ scores, where most 
sparsity-inducing methods presented comparable medians, but ridge 
showed low values with minimal variability due to its design of dense 
solutions. As the network dimension increased, the overall dispersion 
of the performance metrics widened for most methods, reflecting the 
increasing difficulty of high-dimensional precision matrix 
estimation. These results echo the conclusions from 
Table~\ref{tab:sim_res} that MCP and SCAD provide not only strong 
average performance but also robust and stable estimation across 
repeated samples.

\section{Application to TAP Data}
\label{sec:app}

We applied the reviewed precision matrix estimation methods to TAP to 
evaluate their behavior in a clinically relevant resting-state fMRI 
setting. TAP is a well-characterized and diverse clinical cohort 
enrolled by the Vanderbilt Alzheimer’s Disease Research Center, 
consisting of community-dwelling older adults who underwent 
comprehensive clinical, neuropsychological, and neuroimaging 
assessments. Resting-state fMRI (rs-fMRI) was 
collected to characterize functional brain organization and 
connectivity patterns. For the present analysis, we included 
participants with high-quality rs-fMRI data and complete demographic 
and clinical information, resulting in a final sample of $n = 114$ 
participants. Demographic characteristics including age, 
biological gender, \textit{apolipoprotein E4} 
(\textit{APOE-$\varepsilon$4}) carrier 
status and race/ethnicity, are summarized in 
Table~\ref{tab:tap_demo}. The goal of this application is to 
construct subject-level functional brain networks and compare the 
performance of precision matrix estimation methods.

\begin{table}[tbp]
	\centering
	\caption{Demographics for the analysis sample ($n = 114$). 
	Continuous variables are summarized as mean (standard deviation), 
	and categorical variables are presented as counts (percentages). 
	}
	\label{tab:tap_demo}
	\begin{tabular*}{\textwidth}{@{\extracolsep{\fill}}lccccc}
		\toprule
		 & Overall & NC & CI (not MCI) & MCI & Dementia \\
		\midrule
		Sample size & 114 & 41 & 7 & 45 & 21 \\
		Age (Years) &  71.4 (6.6) & 69.0 (6.5) & 72.4 (6.4) & 72.0 
		(6.5) & 74.3 (6.1) \\ 
		Gender (F, \%) & 58.8 & 53.7 & 42.9 & 62.2 & 66.7 \\
		\textit{APOE-$\varepsilon$4} (Pos, \%) & 49.1 & 39.0 & 42.9 & 
		51.1 & 
		66.7 \\
		R/E (NHW, \%) & 65.8 & 56.1 & 57.1 & 71.1 & 76.2 \\
		\bottomrule
	\end{tabular*}
	
	\vspace{2mm}
	\begin{minipage}{\linewidth}
		\footnotesize
		NC: normal cognition; CI: cognitive impairment; MCI: mild 
		cognitive impairment; \textit{APOE-$\varepsilon 4$}: 
		\textit{apolipoprotein E4 allele}; R/E: race and ethnicity; 
		NHW: non-Hispanic white.
	\end{minipage}
	\end{table}

TAP participants underwent multi-modal brain MRI on a 3T Philips 
Achieva system (Best, the Netherlands) with a 32-channel 
phased-array SENSE receiver head coil. For co-registration, 
T1-weighted MPRAGE images were acquired using the following 
parameters: $\text{TR} = 6.5 \text{ms}$, $\text{TE} = 2.9 
\text{ms}$, and $\text{spatial resolution} = 1 \times 1 \times 1 
\text{mm}^3$. Resting-state BOLD-fMRI data were acquired with 
$\text{TR} = 2{,}000 \text{ms}$, $\text{TE} = 30 \text{ms}$ and 
$\text{reconstructed voxel size} = 1.65 \times 1.65 \times 3.00 
\text{mm}^3$ with acquisition length of $6$ minutes yielding $n = 
180$ time points. Resting-state BOLD-fMRI data were then 
pre-processed using established procedures. Data first underwent 
despiking, slice timing correction, and motion 
correction~\citep{jo2013effective}. Images were corrected for 
geometric distortions using 
SynBOLD-DisCo~\citep{yu2023synbolddisco}. Motion artifacts were 
identified and regressed from data using ICA-AROMA through the 
aggressive denoising option~\citep{pruim2015evaluation}. Finally, 
data were co-registered to T1 weighted images and parcellated into 
$p = 400$ ROIs using the Schaefer local-global 
parcellation~\citep{schaefer2017localglobal}. The mean BOLD 
signal for each ROI was then extracted using the \pkg{nilearn} 
package~\citep{pkg:nilearn}. The resulting ROI-by-time matrix for 
each subject was used 
as the input for precision matrix estimation, where the 
parameter tuning was performed by default through 5-fold CV.

\begin{figure}[tbp]
	\centering
	\label{fig:heatmap}
	\includegraphics[width=\textwidth]{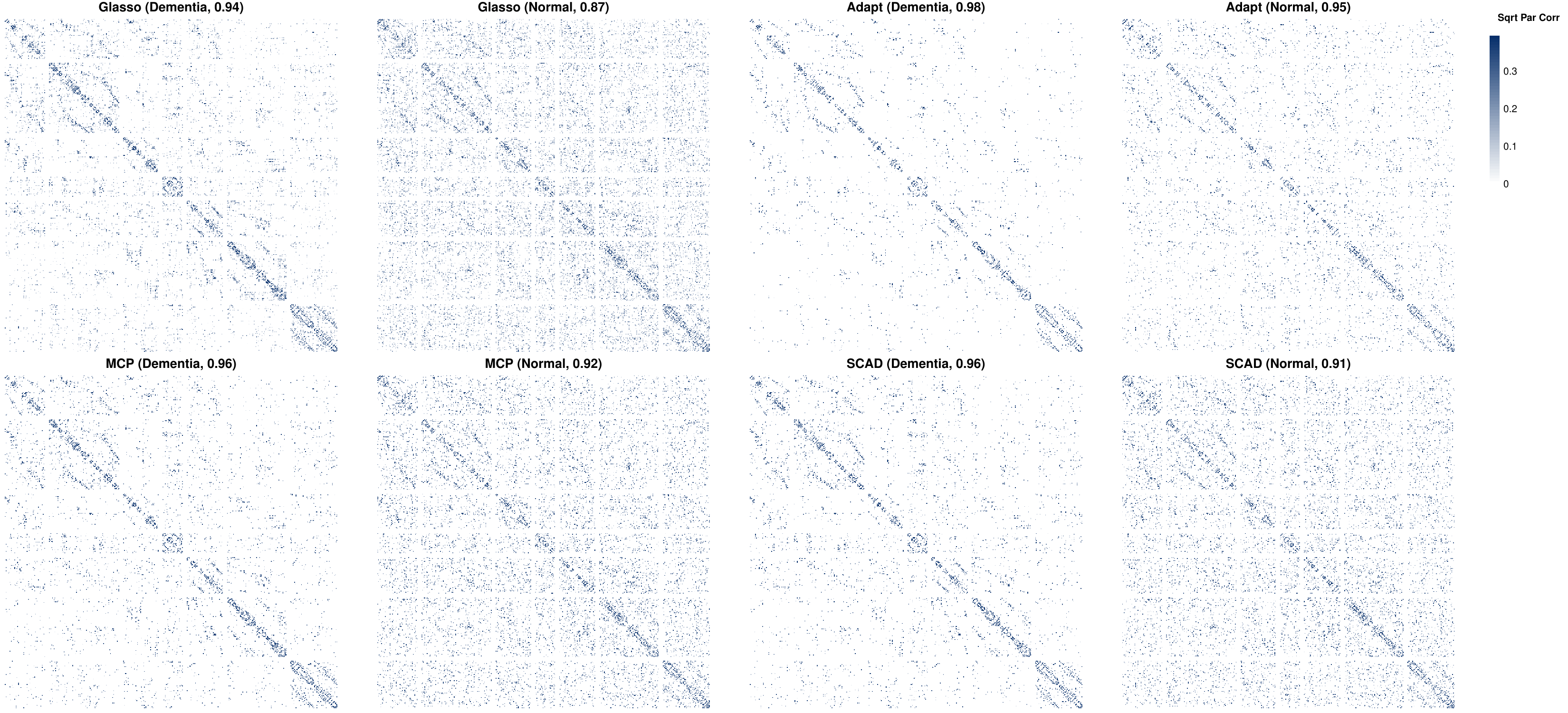}
	\caption{Partial-correlation-based functional connectivity (with 
	$p = 400$) for two participants (NC and dementia) based on 
	estimated precision matrices using glasso, adaptive glasso, MCP, 
	and SCAD. Negative partial correlations were excluded, and a 
	square-root transformation was applied to improve visualization 
	contrast. The sparsity is indicated in the subtitle of each 
	panel.}
\end{figure}

Figure~\ref{fig:heatmap} shows the partial-correlation-based 
functional connectivity for two TAP participants, one with NC and one 
with dementia, obtained by transforming the estimated precision 
matrices. Across all methods and both subjects, the connectivity 
patterns exhibit a clear modular organization, characterized by 
block-like structures along the diagonal. This consistent large-scale 
organization suggests that the overall functional architecture is 
well characterized across estimators, indicating robustness of the 
major network topology to the choice of precision matrix estimation 
method. Comparing the two subjects, the cognitively normal 
participant generally shows stronger and denser connectivity 
patterns, whereas the dementia participant exhibits weaker and more 
fragmented connections, consistent with reduced functional 
integration.

Method-specific differences are primarily reflected in the degree of 
sparsity and the distribution of connection strengths. Adaptive 
glasso produces relatively sparse networks for both subjects, 
retaining only a small number of strong connections while 
aggressively shrinking weaker associations. MCP and SCAD yield 
intermediate sparsity, preserving moderate-strength connections while 
still enforcing substantial regularization. However, for the dementia 
subject, the SCAD estimate appears particularly sparse. In contrast, 
the standard glasso tends to produce denser matrices, reflecting 
relatively uniform shrinkage. These visual differences illustrate the 
distinct regularization behavior across methods and demonstrate that 
the choice of estimator can meaningfully influence the network 
density and connectivity strength. It is important to note that these 
observations are based on individual subjects and are intended to 
provide qualitative illustration of methodological differences rather 
than formal group-level inference.

Additionally, we examined whether the number of hub regions differed 
across diagnostic groups (NC, CI but not MCI, MCI, and dementia). In 
particular, we considered betweenness centrality, a widely used 
measure for identifying hub regions in functional brain 
networks~\citep{kundu2019anovel, oldham2019thedevelopment, 
stam2024hub}. Betweenness centrality quantifies the extent to which a 
node lies on shortest paths between other 
nodes~\citep{brandes2001afaster, xiao2022incorporating}, and thus 
reflects its role in facilitating information transformation 
across brain systems. Regions with high betweenness centrality are 
often interpreted as connector hubs that support large-scale 
communication and network efficiency. For each subject, partial 
correlation matrices derived from the estimated precision matrices 
were first restricted to positive connections by setting negative 
values to zero, and diagonal entries were removed. Different from 
\citet{kundu2019anovel}, who analyzed binarized connectivity 
networks, we computed weighted betweenness 
centrality~\citep{brandes2001afaster}, where 
connection strength was treated as the inverse of edge length when 
determining shortest paths. Hub regions were defined as nodes with 
standardized betweenness values exceeding a threshold (i.e., $z > 
2$), and the total number of hubs (glasso: $16.07 \pm 2.79$; adapt 
glasso: $15.06 \pm 2.75$; MCP: $15.98 \pm 2.42$; SCAD: $15.76 \pm 
2.44$) was used as a summary measure of network organization.

To assess group differences, we conducted one-way analyses of 
variance (ANOVA) across diagnostic groups for each precision matrix 
estimation method. For networks estimated using the standard lasso 
and SCAD, no significant differences in hub number were observed 
across diagnostic groups (glasso: $F = 0.507$, $p = 
0.678$; SCAD: $F = 1.161$, $p = 0.328$). The adaptive glasso showed a 
trend toward group differences, although the overall effect did not 
reach conventional significance levels ($F = 2.326$, $p = 0.079$). In 
contrast, hub counts derived from the MCP-based functional brain 
networks exhibited a statistically significant group effect ($F = 
3.217$, $p = 0.026$), suggesting that network organization estimated 
by MCP may be more sensitive to clinical heterogeneity. Post-hoc 
Tukey comparisons indicated that the difference was primarily driven 
by the contrast between the MCI and dementia groups, with the 
dementia group showing significantly fewer hubs than the MCI group 
($p_{\text{adj}} = 0.014$), where the p-value was adjusted for 
multiple comparisons. No other pairwise comparisons reached 
statistical significance.

Because demographic and genetic factors such as age, biological 
gender, \textit{APOE-$\varepsilon$4}, and race/ethnicity may 
influence both 
brain network organization and cognitive status, we further evaluated 
diagnostic group differences in hub number using analysis of 
covariance (ANCOVA), adjusting for age, biological gender, 
\textit{APOE-$\varepsilon$4} carrier status, and race/ethnicity 
aiming to 
assess whether differences in network topology across diagnostic 
groups persist after accounting for potential confounding effects. 
The ANCOVA suggested an overall effect of diagnosis on hub number 
($F = 2.718$, $p = 0.048$, while none of the covariates showed 
significant associations with hub count. Post-hoc pairwise 
comparisons with p-value adjustment once again showed that the 
difference between the MCI and dementia groups was statistically 
significant ($p_{\text{adj}} = 0.035$), whereas other contrasts were 
not significant. ANCOVA applied to the hub counts derived from the 
adaptive glasso, adjusting for the same set of covariates, did not 
lead to a significant overall effect of diagnosis ($F = 1.883$, $p = 
0.137$). This finding suggests that, after accounting for demographic 
and genetic factors, diagnostic group differences in hub number were 
not evident for networks estimated using the adaptive glasso.

These findings indicate modest but detectable differences in 
functional network organization across diagnostic groups when 
precision matrices were estimated using MCP. Compared with the other 
approaches, MCP-based betweenness centrality demonstrated greater 
sensitivity to diagnostic-group-related variation, suggesting its 
potential to better capture subtle changes in brain network topology. 
Notably, the group effect for MCP remained significant after 
adjustment for demographic and genetic covariates, indicating that 
the observed differences were not attributed to these factors. In 
summary, MCP appears to complement the standard glasso by offering 
enhanced sensitivity while maintaining comparable interpretability, 
supporting its competitiveness for functional connectivity analysis. 
However, method performance may depend on the choice of network 
topology measures and the characteristics of the data. We do not 
conclude that MCP uniformly outperforms the other methods surveyed in 
this article; rather, these findings highlight its competitive 
performance on this particular application.

\section{Discussion}
\label{sec:dis}

This work provides a comprehensive evaluation of commonly used 
precision matrix estimation methods for functional connectivity 
analysis, together with extensive, data-driven simulations and a 
real-data application from AD research. Across a range of network 
dimensions and sparsity levels, the synthetic data analysis results 
demonstrate that different regularization methods lead to differences 
in estimation accuracy, sparsity, computational efficiency, and 
downstream network topology. In particular, MCP and SCAD have shown
competitive performance relative to standard approaches like glasso, 
while adaptive and convex methods have exhibited distinct sparsity 
and shrinkage behaviors. The real-data analysis further illustrates 
that methodological choice can influence the identification of 
network topology and the sensitivity of topology-based clinical 
comparisons. These findings provide practical guidance for 
researchers selecting precision matrix estimators and highlight the 
importance of considering methodological variability when 
interpreting functional connectivity results.

Several important directions remain for future research. First, the 
Gaussian graphical models considered here represent undirected, 
conditional dependence structures and do not distinguish causal or 
directional interactions. Extending functional connectivity analysis 
to directed frameworks would allow investigation of causal influences 
and the potential presence of collider structures, which may play an 
important role in brain network dynamics. Second, current 
subject-level estimation approaches assume homogeneous network 
structures, whereas substantial between-subject heterogeneity often 
exists in clinical populations. Methods that explicitly model 
group-level structure, subgroup heterogeneity, or hierarchical 
variability may improve both estimation stability and 
interpretability. Finally, integrating precision-matrix-based 
connectivity with causal modeling frameworks and longitudinal designs 
may provide deeper insight into disease-related network alterations 
and their temporal evolution. Advances along these directions will 
further enhance the interpretability and clinical utility of 
network-based neuroimaging analyses.

\section*{Acknowledgment}

TAP is funded by NIA (P20-AG068082 and P30-AG086403). JY was 
partially supported by the NSF grant DMS2210735 and a seed grant 
from the Connecticut Institute for the Brain and Cognitive Science, 
University of Connecticut (UConn).

The computational work for this project was conducted using 
resources provided by the UConn Storrs High-Performance Computing 
(HPC) cluster. We extend our gratitude to the UConn Storrs HPC and 
its team for their resources and support, which aided in achieving 
these results.

\bibliographystyle{chicago}
\bibliography{refs}
	
\end{document}